\documentclass[12pt]{article}
\usepackage[T2A]{fontenc}
\usepackage[utf8]{inputenc}
\usepackage[english]{babel}
\usepackage{amsmath, amssymb, amsfonts, amsthm, biblatex}
\usepackage{graphicx}
\usepackage[unicode, pdftex]{hyperref}
\usepackage[left=3cm, top=2cm, right=1cm, bottom=2cm]{geometry}
\fontsize{14pt}{21pt}
\usepackage{setspace}
\numberwithin{equation}{section}

\usepackage[tableposition=top]{caption}
\usepackage{subcaption}
\usepackage{ragged2e}
\title{Scalar field in Bianchi type-I cosmology with Lyra’s
	geometry}
\author{Evgeny D. Petuhov ${}^{1, \,2}$\footnote{e-mail: ped.19@uni-dubna.ru}, Bijan Saha ${}^{1, \,2, \, 3}$\footnote{e-mail: bijan@jinr.ru} }
\date{}

\begin{document}
	\maketitle
	\thispagestyle{empty}
	\begin{flushleft}
		{\it ${}^{1}$ Dubna State University, 141980 Dubna, Moscow Region, Russia\\
			${}^{2}$ Laboratory of Information Technologies, Joint Institute for Nuclear Research, 141980 Dubna, Moscow region, Russia\\
			${}^{3}$ Peoples’ Friendship University of Russia (RUDN University), 117198 Moscow, Russia}
	\end{flushleft}
	\begin{abstract}
		\begin{doublespace}
			\par  In this study, we examine the role of a scalar field in the evolution of the Universe within the framework of a Bianchi type-I
			cosmological model with Lyra’s geometry. Previous research has explored the nonlinear spinor field in various anisotropic and isotropic cosmological
			models. In our current study, we find dynamical restrictions for Lyra parameters and violation of stress-energy tensor conservation within Lyra geometry. We shown that in considering cases behavior of Lyra's parameter corresponds to relative influence in early universe and absence of Lyra geometry in present universe.
		\end{doublespace}
	\end{abstract}
	\newpage

	\setcounter{footnote}{1}
	\setcounter{equation}{1}
	\newpage
	
	\section{Introduction}\label{sec:introduction}
	\par This work dedicated to study modified Riemann geometry in context of cosmological models. Originally, after Einstein’s ground breaking work, Weyl sought to unify gravity and the electromagnetic field by generalizing Riemannian geometry \cite{1}. However, Weyl’s theory was not widely accepted, as it contradicted several well-established observational results. In 1951, Lyra proposed
	a modification of Riemannian geometry that closely resembles Weyl’s theory\cite{2}. Lyra’s approach maintains a metric-preserving connection, similar to that of Riemannian geometry. He introduced a gauge function into the otherwise
	structure-less manifold. Lyra’s theory was further developed by Sen \cite{3}, Halford \cite{4}, Sen and Dunn \cite{5}, Sen and Vanstone \cite{6}, and many others. In \cite{3}, Sen considered a static spherically symmetric cosmological model based on Lyra’s geometry.
	\par After the discovery and further confirmation of late time accelerated mode of expansion \cite{7,8,9} different attempts are being made to explain
	this new found phenomenon. Several approaches have been proposed to solve this problem. While some researchers introduced a repulsive force such as cosmological constant \cite{10,11}, quintessence \cite{12,13}, Chaplygin gas \cite{14,15}, etc., known as dark energy in the R.H.S. of Einstein system of equations, others changed the geometry of space-time introducing higher order invariants of gravitational field giving rise to $f(R)$ \cite{16}, $f(R,\,T)$ \cite{17}, Gauss-Bonnet theory \cite{18}, etc. Lyra’s geometry was exploited by different authors \cite{19,20,21,22}. Cosmological models in Lyra’s geometry for Bianchi type-I space-time were considered in \cite{23} and \cite{24}.
	
	\par In his work Lyra introduced a scalar function in order to modify the both Riemann and Weyl geometry. On the other hand Brans and Dicke
	has introduced a non-minimally coupled scalar and gravitational field that leads to the modification of Einstein gravity \cite{25}.
	Given the similarity Faraoni made an attempt to connect Lyra's geometry with Brans-Dicke theory \cite{26}, which in our view needs a more detailed analysis. Description of dark energy and accelerating expansion of universe in Lyra's geometry background was tested in \cite{27}. Scalar field in Lyra geometry also was studied in work \cite{28}. But scalar field was used like source in right side of Einstein equation, excluding equation of motion of scalar field. Also scalar field used in cosmological inflation models \cite{29} and for modeling dark energy (K-essence models) \cite{30}. In our study we taking to account equation of motion and conservation of energy momentum tensor. We consider naive scenario when scalar field takes in standard form without influence of Lyra's geometry. In future works we hope to construct scalar field theory with Lyra's geometry aspects. Recently, role of nonlinear spinor field in the evolution of the BI universe with Lyra geometry was studied in \cite{31}. In this paper we extend that duty for scalar field.
	
	\section{Geometrical definitions}\label{sec:geometrical_definitions}
Modification of gauge in Weyl geometry by structure-less function is the point of Lyra geometry. According to Lyra’s geometry the displacement vector from a point $P(x^{\mu})$ to a neighboring point $P^{\prime}(x^{\mu} + dx^{\mu})$ is defined by $\xi^{\mu} = x^0 dx^{\mu}$, where $x^0$ is a nonzero analytical function of coordinates and fixes the gauge of the system. Together with coordinate system $x^{\mu}$, $x^0$ form a so-called reference system $(x^0,\, x^{\mu})$. The transformation to a new reference system is given by
\begin{equation}\label{coordinate transform}
	x^{\mu} = x^{\mu}(\bar{x}^1, \dots , \bar{x}^n),\hspace{1cm} x^0 = x^0(\bar{x}^1, \dots , \bar{x}^n, \bar{x}^0),
\end{equation}
where $\partial x^0 /\partial\bar{x}^0 \neq 0 $ and $\text{det} (\partial x^{\mu}/\partial\bar{x}^{\nu}) \neq 0$. Under the transformation (\ref{coordinate transform}) a contravariant vector $\xi^{\mu}$ is transformed according to
\begin{equation}
	\bar{\xi}^{\mu} = \lambda A^{\mu}_{\nu} \xi^{\nu}, \hspace{1cm} A^{\mu}_{\nu} = \frac{\partial\bar{x}^{\mu}}{\partial x^{\nu}}, \hspace{1cm} \lambda = \frac{\bar{x}^0}{x^0}
\end{equation}
with $\lambda$ being the gauge factor of transformation.

In any general reference system $(x^0, \, x^{\mu})$ the infinitesimal parallel transfer of a vector from $P(x^{\mu})$ to $P^{\prime}(x^{\mu} + dx^{\mu})$ can be expressed as
\begin{equation}\label{parallel transfer}
	\delta \xi^{\mu} = -\tilde{\Gamma}^{\mu}_{\alpha\beta} \xi^{\alpha} x^0 dx^{\beta}, \hspace{1cm} \tilde{\Gamma}^{\mu}_{\alpha\beta} = \bar{\Gamma}^{\mu}_{\alpha\beta} -\frac12 \delta^{\mu}_{\alpha} \phi_{\beta}, \hspace{1cm} \phi_{\alpha} = -\frac{1}{x^0}\frac{\partial (\ln \lambda^2)}{\partial x^{\alpha}}.
\end{equation}
It should be noted that $\bar{\Gamma}^{\mu}_{\alpha\beta} = \bar{\Gamma}^{\mu}_{\beta\alpha}$, though $\tilde{\Gamma}^{\mu}_{\alpha\beta}\neq \tilde{\Gamma}^{\mu}_{\beta\alpha}$.

Since the displacement vector between two neighboring points $P(x^{\mu})$ and $P^{\prime}(x^{\mu} + dx^{\mu})$ in this case is define by $\xi^{\mu} = x^0dx^{\mu}$, the interval between them is given by the invariant
\begin{equation}
	ds^2 = g_{\mu\nu}x^0 dx^{\mu} x^0 dx^{\nu},
\end{equation}
where $g_{\mu\nu}$ is the symmetric tensor of second rank. The parallel transport of length in Lyra geometry is integrable, i.e., $\delta(g_{\mu\nu}\xi^{\mu}\xi^{\nu}) = 0$ and the connection $\bar{\Gamma}^{\alpha}_{\mu\nu}$ in (\ref{parallel transfer}) takes form
\begin{equation}\label{Weyl}
	\bar{\Gamma}^{\alpha}_{\mu\nu} = \frac{1}{x^0}\Gamma^{\alpha}_{\mu\nu} + \frac12 \left(\delta^{\alpha}_{\mu}\phi_{\nu} + \delta^{\alpha}_{\nu}\phi_{\mu} - g_{\mu\nu}\phi^{\alpha} \right),
\end{equation}
which is similar to that of Weyl geometry except the multiplier $1/x^0$. Here $\Gamma^{\alpha}_{\mu\nu}$ is the Levi-Civita connection. Note that in Lyra geometry the derivative $\partial/\partial x^{\mu}$ is substitute by $\partial/(x^0\partial x^{\mu}) = (1/x^0)\partial/\partial x^{\mu}$. From \eqref{Weyl} and \eqref{parallel transfer} we find the connection for Lyra geometry:
\begin{equation}\label{LyraCon}
	\tilde{\Gamma}^{\alpha}_{\mu\nu} = \frac{1}{x^0}\Gamma^{\alpha}_{\mu\nu} + \frac12 g^{\alpha \tau} \left( g_{\nu\tau}\phi_{\mu} - g_{\nu\mu}\phi_{\tau} \right),
\end{equation}
which explicitly shows the presence of torsion.

Einstein’s field equation in Lyra’s geometry in normal gauge $(x^0 = 1)$ was found by Sen \cite{3}:
\begin{equation}\label{einstein equation}
	G^{\nu}_{\mu} + \frac32 \phi_{\mu}\phi^{\nu} -\frac34 \delta^{\nu}_{\mu} \phi_{\alpha} \phi^{\alpha} = \kappa T^{\nu}_{\mu},
\end{equation}
where $\phi_\mu$ is the displacement vector. Taking 4-divergence of (\ref{einstein equation}), on account of Bianchi identity $G^{\nu}_{\mu ; \nu} = 0$, we find the equation for displacement vector $\phi_\alpha$:
\begin{equation}\label{bianchi condition}
	\left( \frac32 \phi_{\mu}\phi^{\nu} -\frac34 \delta^{\nu}_{\mu} \phi_{\alpha} \phi^{\alpha} \right)_{;\nu} = T^{\nu}_{\mu ; \nu}.
\end{equation}
We exploit this equation to find displacement vector $\phi_\alpha$.

\section{Basic equations}\label{sec:basic_equations}

Let us now study the scalar field within the scope of Bianchi type-I space-time. The metric is given by
\begin{equation}\label{3d metric}
	ds^2 = dt^2 - a^2_1 dx^2_1 - a^2_2 dx^2_2 - a^2_3 dx^2_3 ,
\end{equation}
where $a_1, a_2, a_3$ -- only time depended functions and in general $a_1 \ne a_2 \ne a_3$. This is a strait-forward generalization of FRW universe.

Scalar field in our cosmological model is given by the Lagrangian
\begin{equation}
	\mathcal{L} = \frac{1}{2} g^{\mu\nu}\partial_{\mu}\varphi\partial_{\nu}\varphi - U(\varphi).
\end{equation}
Equation of motion for scalar field
\begin{equation}\label{eom for scalar field}
	\frac{1}{\sqrt{-g}} \partial_{\mu}\left( \sqrt{-g}g^{\mu\nu}\partial_{\nu} \varphi\right) + \frac{dU}{d\varphi} = 0.
\end{equation}
We will consider the case when the scalar field depends on time only, i.e., $\varphi = \varphi(t)$. In this case for the scalar field we obtain
\begin{align}
	\ddot \varphi + \left(\frac{\dot{a_1}}{a_1} + \frac{\dot{a_2}}{a_2} + \frac{\dot{a_3}}{a_3} \right) \dot{\varphi} + \frac{dU}{d\varphi} = 0. \label{scfe}
\end{align}

Nontrivial components for stress-energy tensor in case of a time-dependent scalar field are
\begin{eqnarray}
	& T_0^0 = \dfrac12 \dot{\varphi}^2 + U(\varphi), \hspace{2cm} T_1^1 = T_2^2 = T_3^3 = -\dfrac12 \dot{\varphi}^2 + U(\varphi). \label{stress-energy tensor components}
\end{eqnarray}

We consider the case when the displacement vector is given by $ \phi_{\mu} = \left\lbrace \beta (t), \, 0, \, 0, \, 0 \right\rbrace $, with $\beta$ being the parameter of Lyra geometry. Let us examine what happens to energy-momentum tensor (EMT) of scalar field in this case. Inserting \eqref{stress-energy tensor components} into
\begin{align}
	T^{\nu}_{\mu;\nu} &= T^{\nu}_{\mu,\nu} + \tilde{\Gamma}^\nu_{\alpha \nu} T^{\alpha}_{\mu} - \tilde{\Gamma}^\alpha_{\mu \nu} T^{\nu}_{\alpha}
\end{align}
we find
\begin{equation}
	T^{\nu}_{0;\nu} = T^{\nu}_{0,\nu} + \tilde{\Gamma}^\nu_{0 \nu} T^{0}_{0} - \tilde{\Gamma}^\alpha_{0 \nu} T^{\nu}_{\alpha} = \left[ \ddot \varphi + \left(\frac{\dot{a_1}}{a_1} + \frac{\dot{a_2}}{a_2} + \frac{\dot{a_3}}{a_3} \right) \dot{\varphi} + \frac{dU}{d\varphi} \right] \dot\varphi + \frac32 \beta \dot \varphi^2.
\end{equation}
On account of \eqref{scfe} we find
\begin{equation}\label{div from energy tensor}
	T^{\nu}_{0 ; \nu} = \frac32 \beta \dot{\varphi}^2.
\end{equation}

Thus we see that in case of Lyra geometry the energy-momentum tensor of scalar field does not preserve. Then from (\ref{div from energy tensor}) and \eqref{bianchi condition} we obtain the following equation for $\beta$:
\begin{equation}
	\dot{\beta} + \beta \left(\frac{\dot{a_1}}{a_1} + \frac{\dot{a_2}}{a_2} + \frac{\dot{a_3}}{a_3} \right) + \frac32 \beta^2 - \dot{\varphi}^2 = 0. \label{beta}
\end{equation}

The nontrivial components of Einstein system (\ref{einstein equation}) take the form:
\begin{subequations}
	\label{einstein equation in components}
	\begin{eqnarray}
		& \dfrac{\dot{a_1}\dot{a_2}}{a_1 a_2} + \dfrac{\dot{a_2}\dot{a_3}}{a_2 a_3} + \dfrac{\dot{a_1}\dot{a_3}}{a_1 a_3} + \dfrac34 \beta^2 = \kappa \dfrac{\dot{\varphi}^2}{2} + \kappa U, \\
		& \dfrac{\ddot{a_1}}{a_1} + \dfrac{\ddot{a_3}}{a_3} + \dfrac{\dot{a_2}\dot{a_3}}{a_2 a_3} - \dfrac34 \beta^2 = -\kappa \dfrac{\dot{\varphi}^2}{2} + \kappa U, \\
		& \dfrac{\ddot{a_2}}{a_2} + \dfrac{\ddot{a_3}}{a_3} + \dfrac{\dot{a_1}\dot{a_3}}{a_1 a_3} - \dfrac34 \beta^2 = -\kappa \dfrac{\dot{\varphi}^2}{2} + \kappa U, \\
		& \dfrac{\ddot{a_1}}{a_1} + \dfrac{\ddot{a_2}}{a_2} + \dfrac{\dot{a_1}\dot{a_2}}{a_1 a_2} - \dfrac34 \beta^2 = -\kappa \dfrac{\dot{\varphi}^2}{2} + \kappa U.
	\end{eqnarray}
\end{subequations}
Introducing volume scale
\begin{align}
	V &= a_1 a_2 a_3, \label{VolumeScale}
\end{align}
the system (\ref{einstein equation in components}) can be written as \cite{32}:
\begin{equation}
	\frac{\ddot{V}}{V} = 6 \kappa U \label{compact einstein equation}.
\end{equation}

\section{Solutions to the equations}\label{sec:solutions_to_the_equations}

Let us now write the complete system to define metric functions, scalar field and Lyra parameter:
\begin{subequations}
	\label{all system of eom}
	\begin{eqnarray}
		& \ddot{\varphi} + \dfrac{\dot{V}}{V} \dot{\varphi} + \dfrac{dU}{d\varphi} = 0, \label{Scff} \\
		& \dfrac{\ddot{V}}{V} = 6 \kappa U, \label{VSc} \\
		& \dot{\beta} + \dfrac{\dot{V}}{V} \beta + \dfrac32 \beta^2 - \dot{\varphi}^2 = 0.
		\label{LP}
	\end{eqnarray}
\end{subequations}
The equations for scalar field $\varphi$ \eqref{Scff} and volume scale $V$ \eqref{VSc} do not contain $\beta$ explicitly. Thus we solve these equations for a given $U(\varphi)$ \eqref{LP} and further study the evolution of Lyra parameter.

Now let's consider some examples of integrable potentials:
\begin{subequations}
	\begin{eqnarray}
		& U = const, \\
		& U(\varphi) = \exp (\lambda \varphi).
		\label{some potentials}
	\end{eqnarray}
\end{subequations}

\subsection{Case with constant potential}\label{subsec:case_with_constant_potential}

\par System of equation of motion in this case:
\begin{subequations}
	\begin{eqnarray}
		& \ddot{\varphi} + \dfrac{\dot{V}}{V} \dot{\varphi} = 0, \\
		& \dfrac{\ddot{V}}{V} = 6\kappa C, \hspace{1cm} C = const, \\
		& \dot{\beta} + \dfrac{\dot{V}}{V} \beta + \dfrac32 \beta^2 - (\dot{\varphi})^2 = 0.
	\end{eqnarray}
\end{subequations}
Solution is
\begin{subequations}
	\begin{eqnarray}
		& V(t) = V_0 e^{\lambda t} + V_1 e^{-\lambda t}, \hspace{1cm} \lambda \equiv \sqrt{6\kappa C},\\
		& \varphi (t) = \dfrac{\varphi_0}{\sqrt{V_0 V_1}\lambda^2} \arctan\left( \sqrt{\dfrac{V_0}{V_1}}e^{\lambda t} \right) + \varphi_1,
	\end{eqnarray}
\end{subequations}
where $V_0,\, V_1, \, \varphi_0,\, \varphi_1$ -- is integration constants.In the following, similar notations will be used for integration constants in other cases.

\begin{figure}[h!]
	\begin{minipage}[t]{0.45\textwidth}
		\centering
				\includegraphics[width=\textwidth]{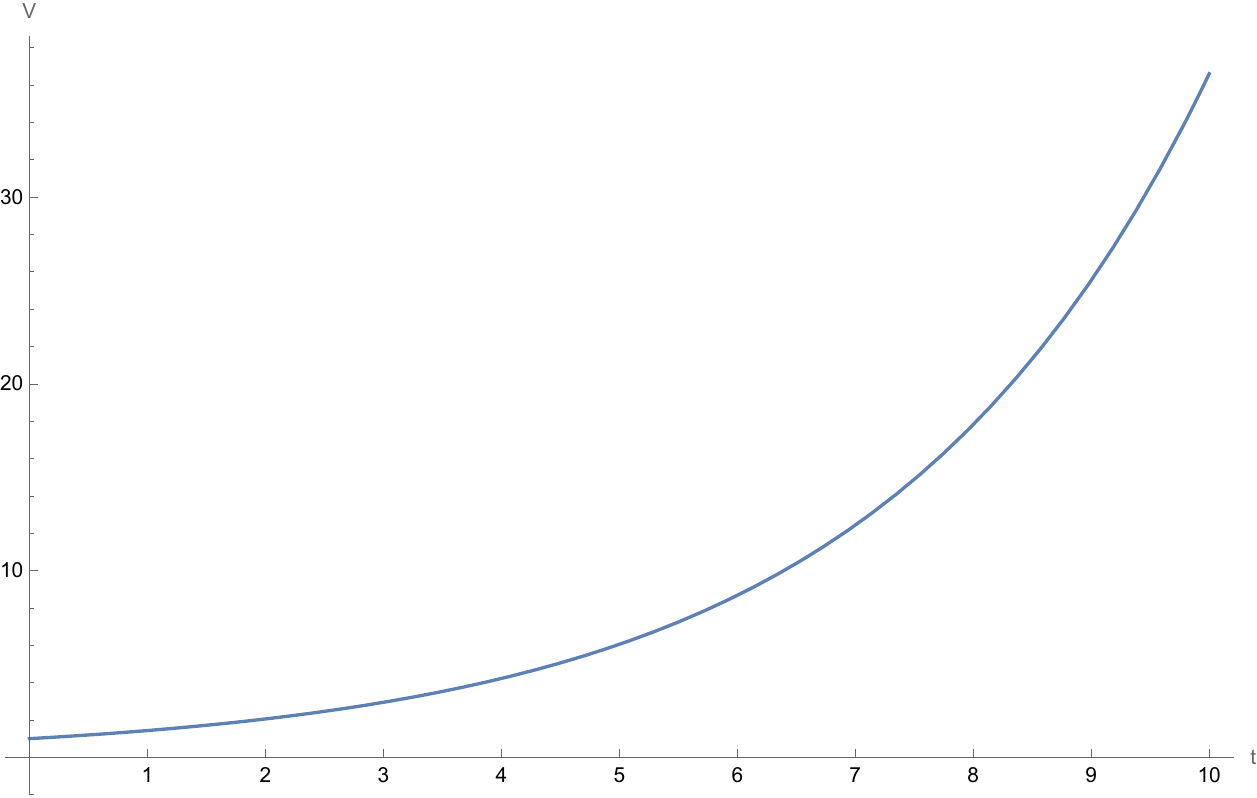}
		\caption{\small Volume function V in case of dark energy}\label{V in dark energy case}
	\end{minipage}
	\begin{minipage}[t]{0.45\textwidth}
		\centering
				\includegraphics[width=\textwidth]{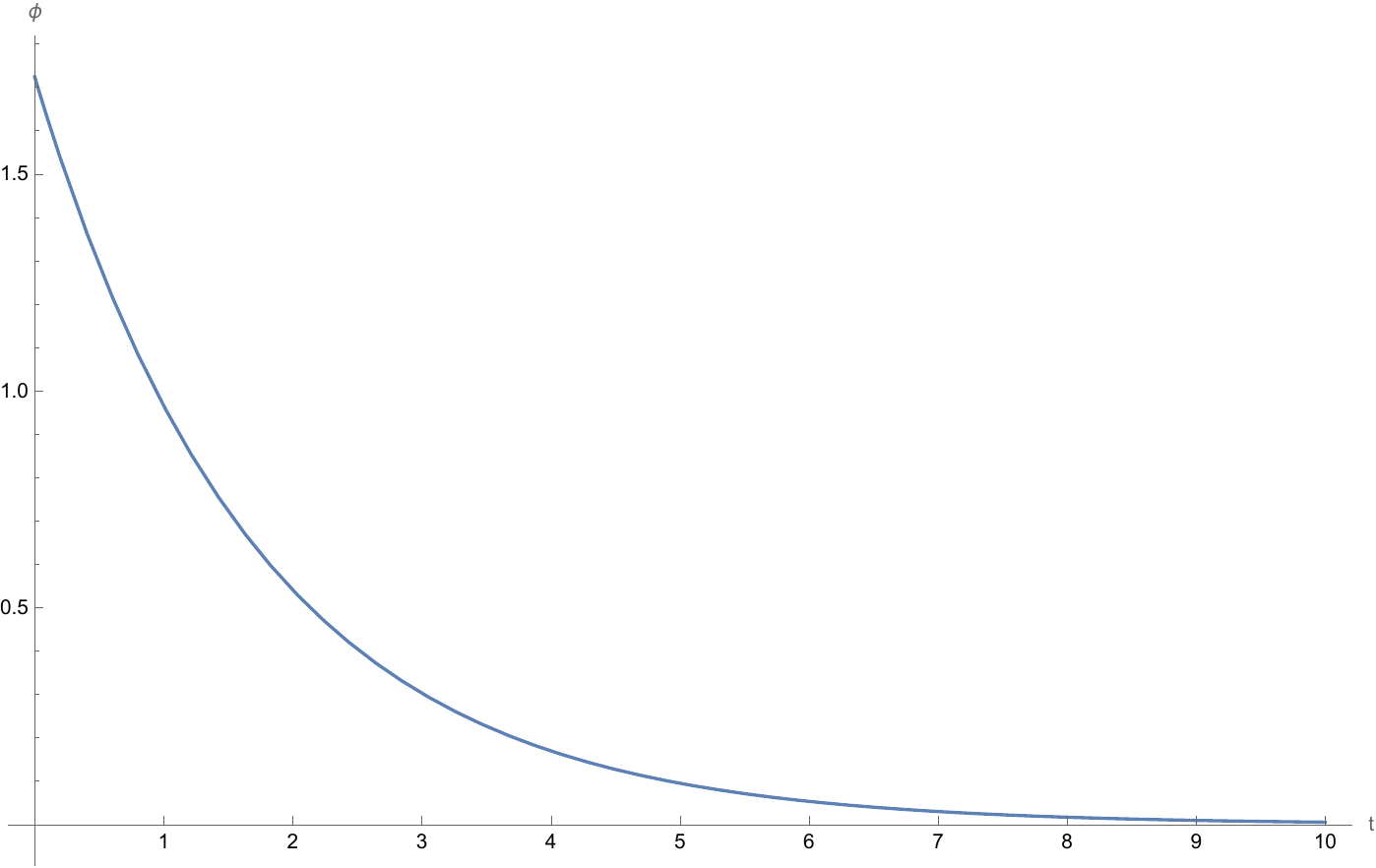}
		\caption{\small Scalar field in case of dark energy}\label{scalar field in case of dark energy}
	\end{minipage}
\end{figure}

We choice private solution when $C_2 = 0$ because universe volume function $V$ needed to describe the accelerating expansion
\begin{subequations}
	\begin{eqnarray}
		& V(t) = V_0 e^{\lambda t}, \hspace{1cm} \lambda \equiv \sqrt{6\kappa C},\\
		& \varphi (t) = \dfrac{\varphi_0}{\lambda}e^{-\lambda t} + \varphi_{\infty}.
	\end{eqnarray}
\end{subequations}

Equation for $\beta$ in this case
\begin{equation}
	\dot{\beta} + \lambda \beta + \frac32 \beta^2 -\varphi_0^2 e^{-2\lambda t} = 0.
\end{equation}

Solution for $\beta$ in this case
\begin{equation}
	\beta (t) = 2 e^{-\lambda t}\sqrt{\frac{-\varphi_0^2}{3}}\tan \left[\frac{\sqrt{-3\varphi_0^2}}{2\lambda} e^{-\lambda t} + \beta_0 \right].
\end{equation}

\begin{figure}[h]
	\center{\resizebox{!}{8cm}{
			\includegraphics{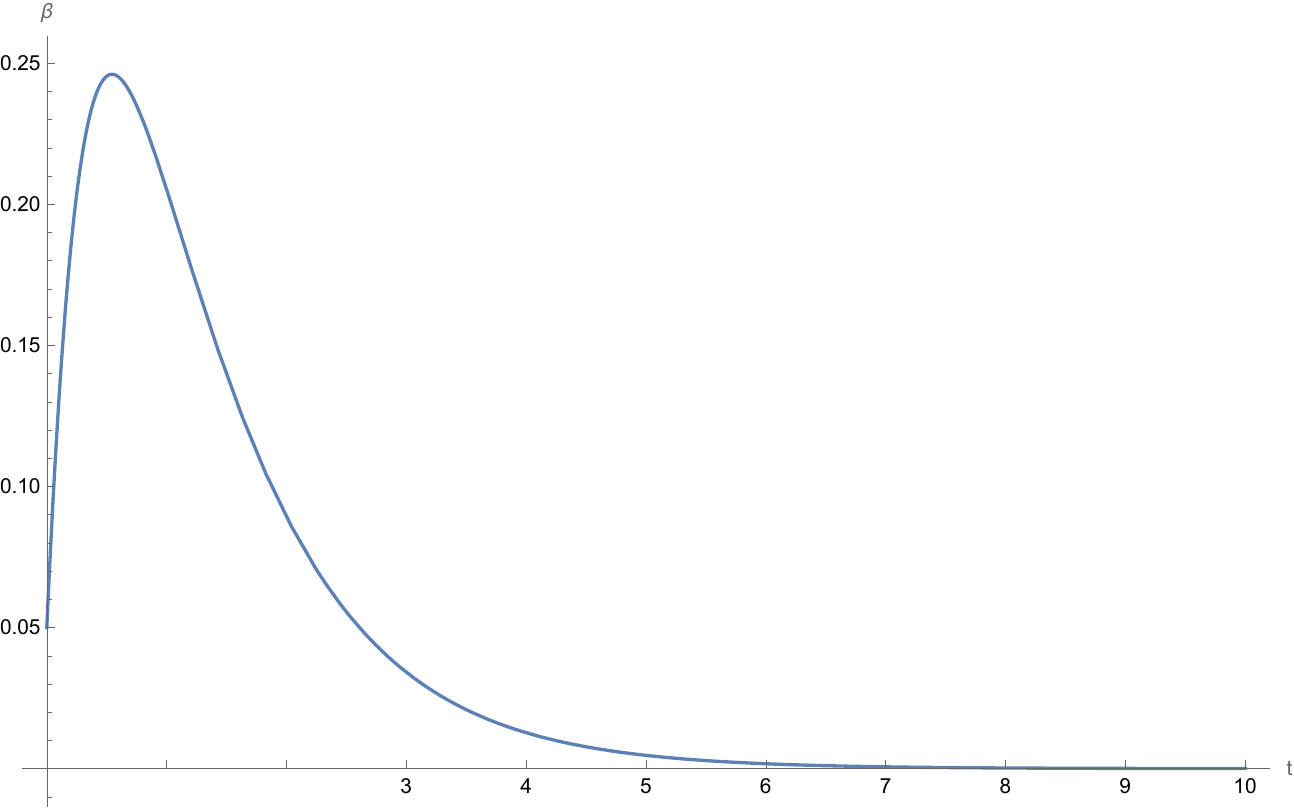}}}
	\caption{\small Lyra's parameter $\beta$ in case of dark energy}\label{beta in case of dark energy}
\end{figure}

The simplest choice for constant potential is trivial one. Solution is
\begin{subequations}
	\begin{eqnarray}
		& V(t) = {V}_0 t + V_1, \\
		& \varphi (t) = {\varphi}_0 \ln\left({V}_0 t + V_1 \right) + \varphi_1, \\
		& 	\beta (t) = -\dfrac{2\sqrt{-V_0^2 \varphi_0^2}}{\sqrt{3}(V_0 t+ V_1)} \tan \left[\dfrac{\sqrt{-3V_0^2 \varphi_0^2}}{2V_0} \ln \left(V_0 t + V_1 \right) + \beta_0 \right],
	\end{eqnarray}
\end{subequations}
where $V_0,\, V_1, \, \varphi_0,\, \varphi_1, \, \beta_0$ -- is integration constants.

\begin{figure}[h!]
	\center{\resizebox{!}{8cm}{
			\includegraphics{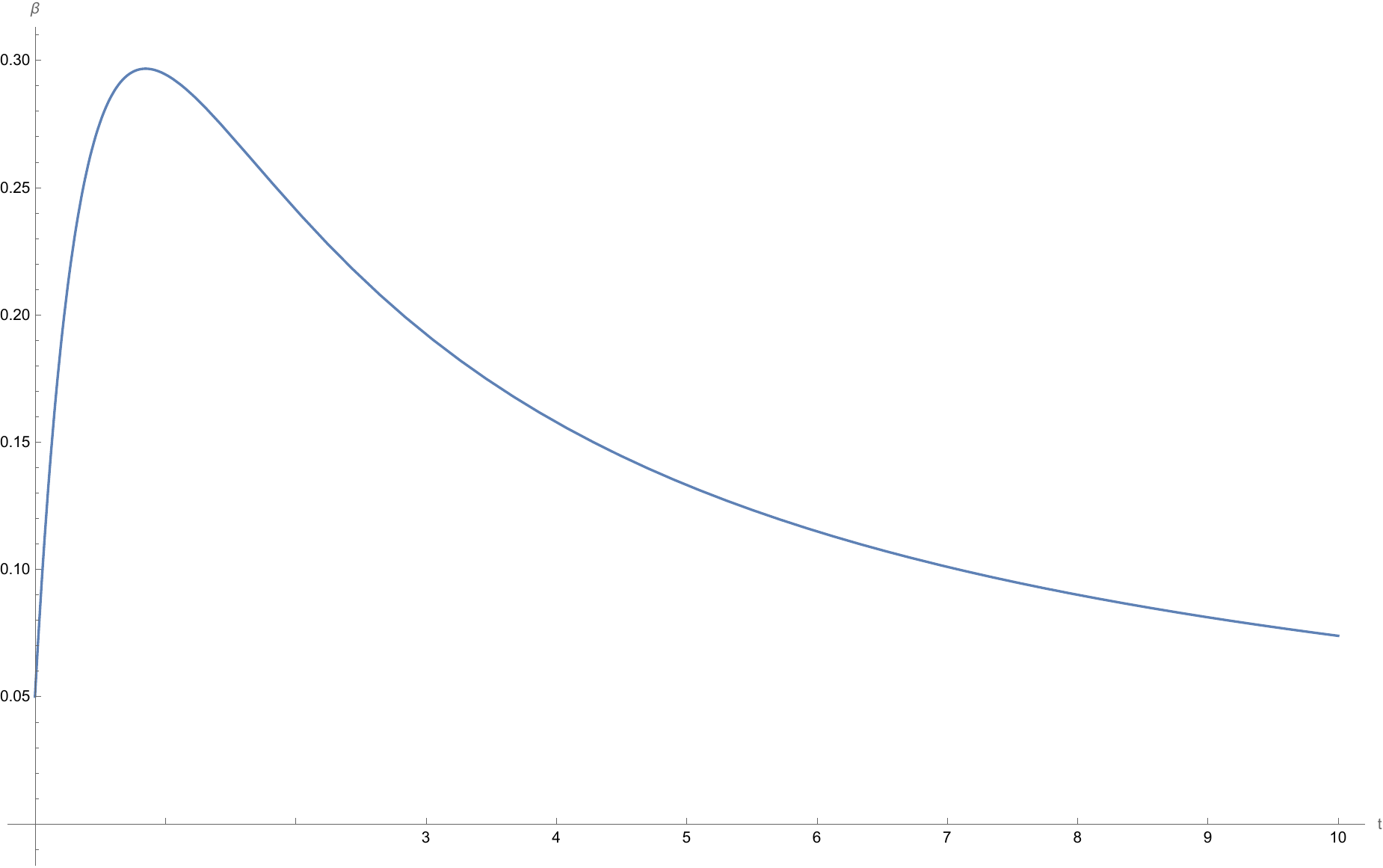}}}
	\caption{\small Lyra's parameter $\beta$ in case of trivial potential}\label{beta in null potential}
\end{figure}

We need to comment this graphic and this comment is actual to other graphics for Lyra's parameter. We used small initial value for $\beta$ in all cases $\beta (0) = 0.05$. When analytical solution is given in terms of complex-value function we used numerical solution and numerical solution as graphic displayed in work.

As we can see, presence of non-zero constant as potential in case Fig. \ref{beta in case of dark energy} leads to more rapid decline of $\beta$ with time then in Fig. \ref{beta in null potential}.

\subsection{Exponential potential}\label{subsec:exponential_potential}

Other example of integrable potentials is exponential potential. System of equation of motion in this case:
\begin{subequations}
	\begin{eqnarray}
		& \ddot{\varphi} + \dfrac{\dot{V}}{V} \dot{\varphi} + \lambda e^{\lambda\varphi} = 0, \\
		& \dfrac{\ddot{V}}{V} = 6\kappa e^{\lambda\varphi}, \\
		& \dot{\beta} + \dfrac{\dot{V}}{V} \beta + \dfrac32 \beta^2 - (\dot{\varphi})^2 = 0.
		\label{eom system for exponent}
	\end{eqnarray}
\end{subequations}
We substitute second equation as third term in first and obtain one equation
\begin{equation}
	\ddot{\varphi} + \dfrac{\dot{V}}{V} \dot{\varphi} = -\frac{\lambda\ddot{V}}{6\kappa V}.
\end{equation}
Multiplying by $V$
\begin{equation}
	V\ddot{\varphi} + \dot{V} \dot{\varphi} = -\frac{\lambda\ddot{V}}{6\kappa}.
\end{equation}
This equivalent to
\begin{equation}
	\frac{d}{dt}(V \dot{\varphi}) = -\frac{\lambda\ddot{V}}{6\kappa}.
\end{equation}
If we consider integration constant as zero we have solution for $\varphi$
\begin{equation}
	\varphi (V(t)) = \ln\left( \varphi_0 V^{-\lambda/6\kappa}\right).
\end{equation}
Back to equation (\ref{eom system for exponent})
\begin{equation}
	\ddot{V} = \frac{6\kappa}{\varphi_0} V^{1-\lambda/6\kappa}.
\end{equation}
If we consider first integration constant as zero we have explicit solution
\begin{equation}
	V (t) = \left(\frac{\lambda^2 \varphi_0}{24\kappa - 2\lambda} \right)^{\dfrac{6\kappa}{\lambda}} (t+t_0)^{\dfrac{12\kappa}{\lambda}}.
\end{equation}
Now we see that for accelerating expansion of universe we need to choice positive $\lambda$. Solution for scalar field
\begin{equation}
	\varphi (t) = \ln \left[\frac{24\kappa - 2\lambda}{\lambda^2} (t+t_0)^{-2} \right].
\end{equation}
Equation for $\beta$
\begin{equation}
	\dot{\beta} + \frac{12\kappa}{\lambda(t+t_0)} \beta + \frac32 \beta^2 -\frac{4}{(t+t_0)^2} = 0.
\end{equation}

\begin{figure}[h!]
	\center{\resizebox{!}{8cm}{
			\includegraphics{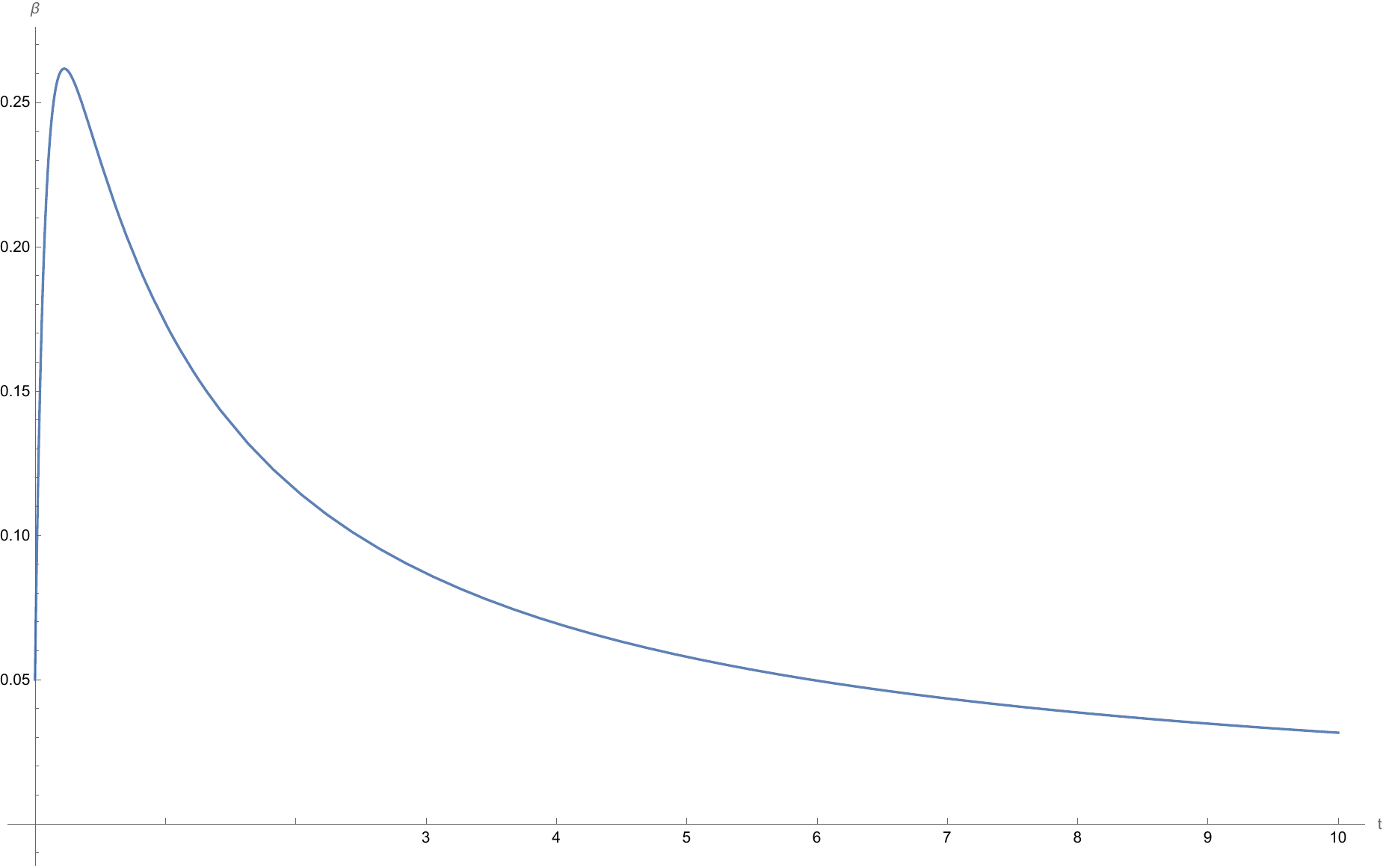}}}
	\caption{\small Lyra's parameter $\beta$ in case of exponential potential}\label{beta in case of exponential potential}
\end{figure}

\section{Equation of motion within equation of state}\label{sec:equation_of_motion_within_equation_of_state}

Equation of state used in cosmology as standard tool for describing different kinds of matter. In sec. 5 we used equation of state as means of simplification of nonlinearity of the potential. Equation of state
\begin{equation}
	W = \frac{p}{\varepsilon} = \frac{-T^1_1}{T^0_0} = \frac{ \dfrac12 (\dot{\varphi})^2 - U(\varphi)}{ \dfrac12 (\dot{\varphi})^2 + U(\varphi)}.
\end{equation}
Potential in terms of $W$ and $\dot{\varphi}$ equal
\begin{equation}
	U = \left(\frac{1-W}{1+W} \right) \frac{(\dot{\varphi})^2}{2} \label{potential in terms W}.
\end{equation}
Let's begin with general case taking into account formula (\ref{potential in terms W}):
\begin{subequations}
	\begin{eqnarray}
		& \dfrac{2V\ddot{\varphi}}{1-W} + \dot{V}\dot{\varphi} = 0, \\
		& \dfrac{\ddot{V}}{V} = 3\kappa \left(\dfrac{1-W}{1+W} \right) \dot{\varphi}^2, \\ \label{eom in general case}
		& \dot{\beta} + \dfrac{\dot{V}}{V} \beta + \dfrac32 \beta^2 - \dot{\varphi}^2= 0.
	\end{eqnarray}
\end{subequations}
Solution for $\dot{\varphi}$
\begin{equation}\label{scalar field anzats}
	\dot{\varphi} = {\varphi_0} V^{\dfrac{W-1}{2}}.
\end{equation}
Substituting (\ref{scalar field anzats}) to equation for $V$ (\ref{eom in general case}) we obtain solution for $V$
\begin{equation}\label{general solution}
	\frac{V}{\sqrt{V_0}}\, {}_2 F_1 \left(\frac12, \, \frac{1}{W}; \, 1+ \frac{1}{W}; \, \frac{6\kappa \varphi_0^2 (W-1)}{{V_0}W(1+W)} V^{1+W} \right) = t+t_0.
\end{equation}
If we consider first integration constant $V_0$ as zero, than we have solution explicitly
\begin{equation}
	V(t) = (-1)^{1/W} \left[\frac{W(1+W)^2}{6\kappa \varphi_0^2 (1-W)} \right]^{1/W} (t+t_0)^{-1/W}.
\end{equation}

Further we consider some cases with fixed $W$. We choice this values of parameter $W$ because in this cases hypergeometric function \eqref{general solution} have straight forward representation in terms of elemental, rational, or polynomial functions.

\subsection{Perfect fluid case}\label{subsec:perfect_fluid_case}

If we consider $W = \dfrac12$ then we have explicit expression for $V$
\begin{equation}
	V(t) = \left[\frac{\varphi_0}{4}\sqrt{\frac{\kappa}{3}} (t+t_0) \right]^4.
\end{equation}
Which leads to explicit solution for scalar field
\begin{equation}
	\varphi (t) = \ln \left[\varphi_1 (t+t_0)^{\sqrt{\dfrac{48}{\kappa}}} \right].
\end{equation}
Equation for $\beta$
\begin{equation}
	\dot{\beta} + \frac{4}{t+t_0} \beta + \frac32 \beta^2 - \frac{48}{\kappa (t+t_0)^2} = 0.
\end{equation}

\begin{figure}[h!]
	\center{\resizebox{!}{8cm}{
			\includegraphics{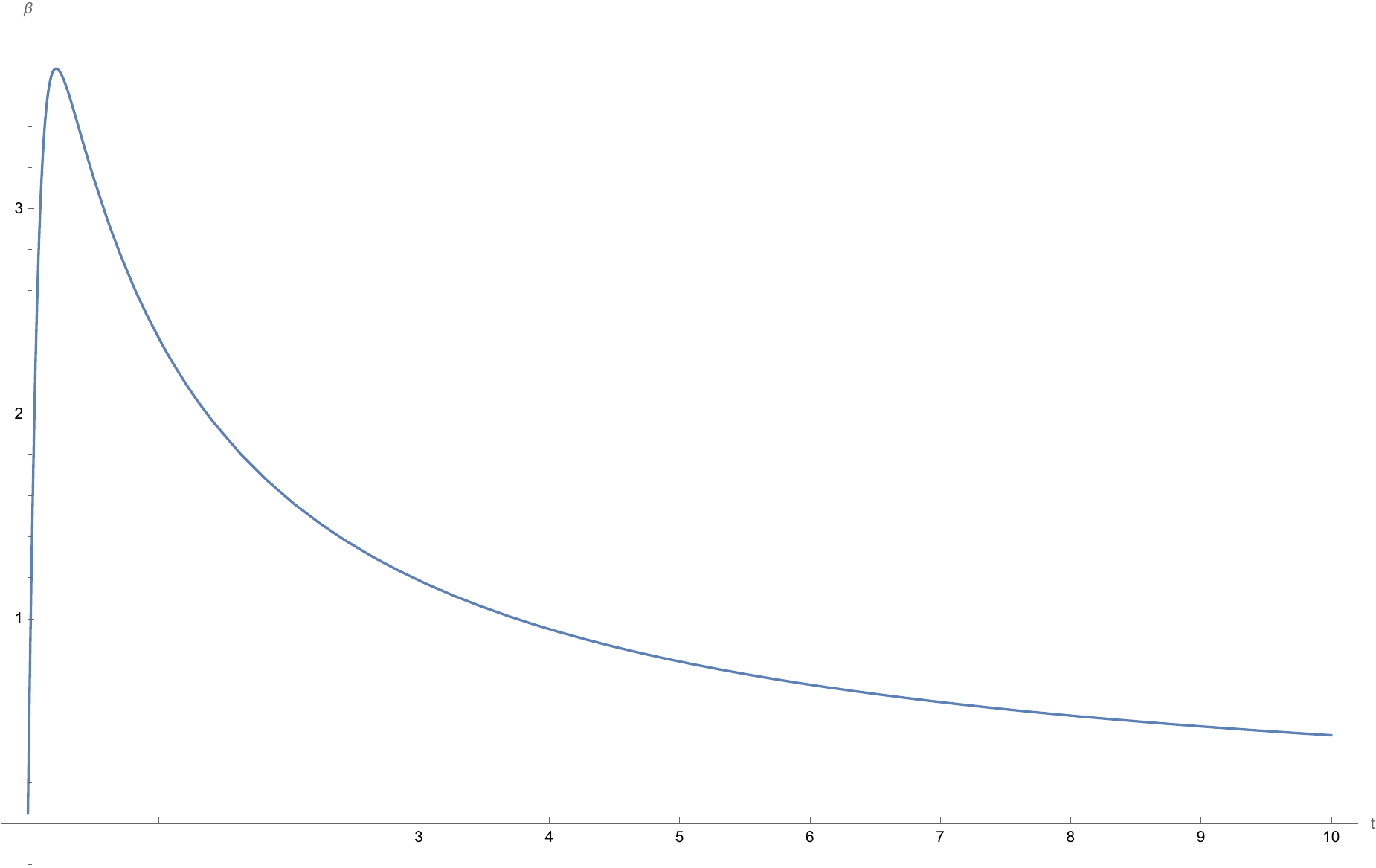}}}
	\caption{\small Lyra's parameter $\beta$ in model with perfect fluid}\label{beta in perfect fluid case}
\end{figure}

\subsection{Exotic matter case}\label{subsec:exotic_matter_case}

If we consider $W = 2$ then we have explicit expression for $V$
\begin{equation}
	V(t) = \frac{1}{\varphi_0}\sqrt{\frac{V_1}{\kappa}}\sin\left[ \sqrt{\kappa V_1} (t+ t_0) \right].
\end{equation}
Which leads to explicit solution for scalar field
\begin{equation}
	\varphi (t) = -\frac{2\varphi_0}{\sqrt{\kappa V_1}} E\left(\frac{\pi}{4} -\frac{\sqrt{\kappa V_1}}{2}(t+t_0), \, 2 \right) + \varphi_1,
\end{equation}
where $E(t,m)$ -- elliptic integral of second kind.

\begin{figure}[h!]
	\center{\resizebox{!}{7cm}{
			\includegraphics{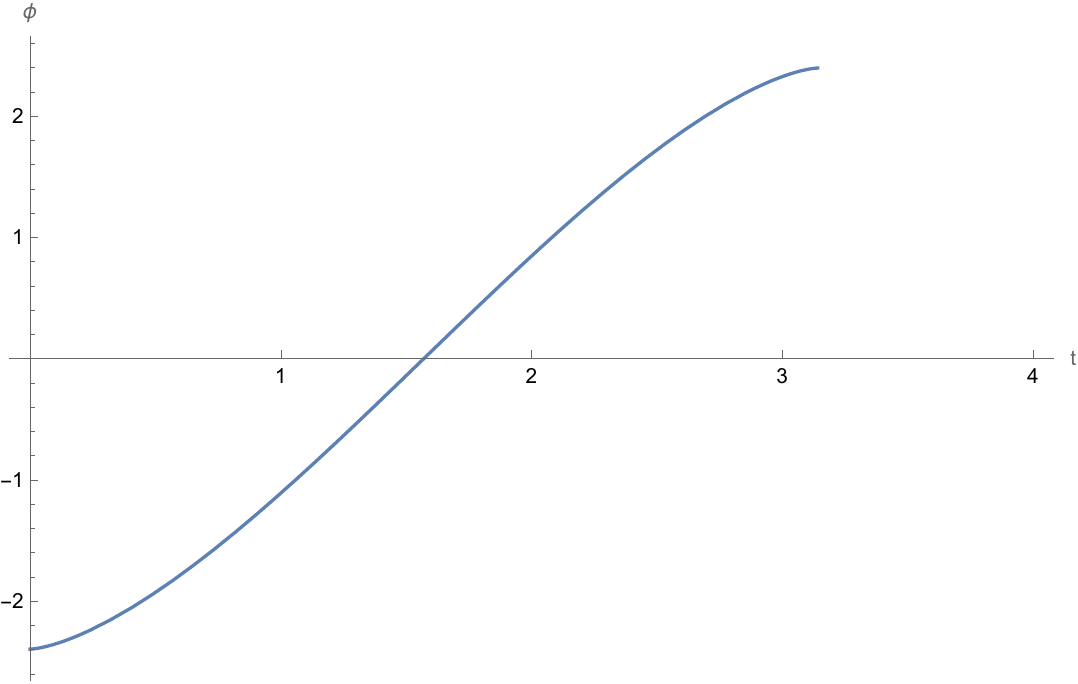}}}
	\caption{\small Scalar field in case W = 2}\label{scalar field in exotic matter}
\end{figure}

Equation for $\beta$
\begin{equation}
	\dot{\beta} + \sqrt{\kappa V_1} \cot \left[\sqrt{\kappa V_1} (t+t_0) \right] \beta + \frac32 \beta^2 - \varphi_0 \sqrt{\frac{V_1}{\kappa}}\sin\left[\sqrt{\kappa V_1} (t+t_0) \right] = 0.
\end{equation}

\begin{figure}[h!]
	\center{\resizebox{!}{7cm}{
			\includegraphics{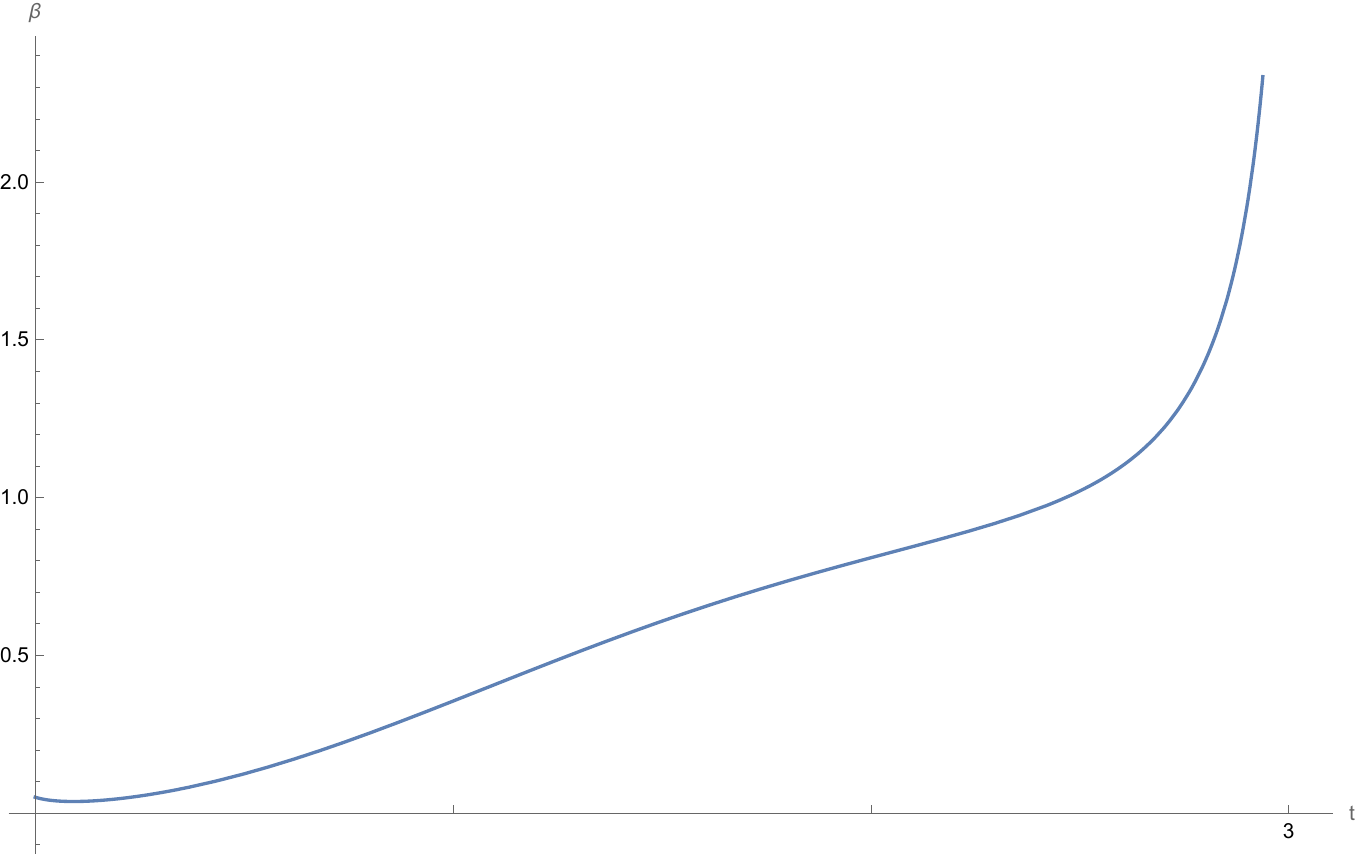}}}
	\caption{\small Lyra's parameter $\beta$ in model with exotic matter (in the vicinity of the point $t=\pi$ $\beta$ is equal infinity because scalar field (\ref{scalar field in exotic matter}) isn't defined)}\label{beta in exotic matter case}
\end{figure}

\subsection{Phantom matter case}\label{subsec:phantom_matter_case}

If we consider $W = - 2$ then we have explicit expression for $V$
\begin{equation}
	V(t) = \sqrt{V^2_1 (t+t_0)^2 - \frac{9\kappa \varphi^2_0}{V_1}}.
\end{equation}
Which leads to explicit solution for scalar field
\begin{equation}
	\varphi (t) = -\frac{(t+t_0)}{9\varphi_0 \kappa}\left[(t+t_0)^2 -\frac{9\varphi_0 \kappa}{V_1^3} \right]^{-1/2}.
\end{equation}

Equation for $\beta$
\begin{equation}
	\dot{\beta} + \frac{V_1^2}{V_1^2 (t+t_0)^2 - \dfrac{9\kappa \varphi_0^2}{V_1}} \beta + \frac32 \beta^2 - \frac{\varphi_0^2}{\left( V_1^2 (t+t_0)^2 - \dfrac{9\kappa \varphi_0^2}{V_1}\right) ^{3/2}} = 0.
\end{equation}

\begin{figure}[h!]
	\center{\resizebox{!}{7.5cm}{
			\includegraphics{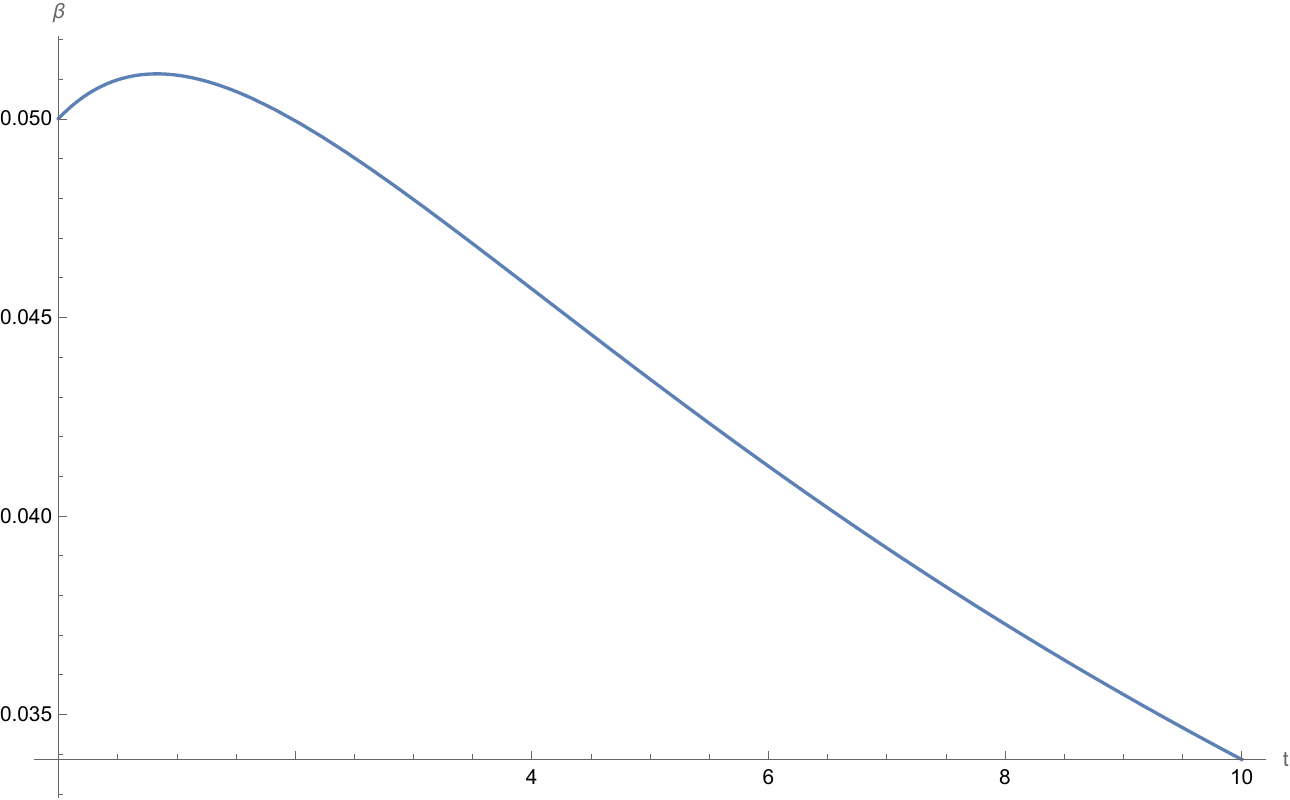}}}
	\caption{\small Lyra's parameter $\beta$ in model with phantom matter}\label{beta in plantom matter case}
\end{figure}

\subsection{Quintessence model}\label{subsec:quintessence_model}

If we consider $W = - \dfrac12$ then we have explicit expression for $V$
\begin{equation}
	V(t) = \left[\frac{9\varphi_0}{2} \sqrt{\frac{\kappa}{2}} (t+t_0) \right]^{\dfrac43}.
\end{equation}
Which leads to explicit solution for scalar field
\begin{equation}
	\varphi (t) = \ln \left[\varphi_1 (t+t_0)^{\sqrt{\dfrac{8}{81\kappa}}} \right].
\end{equation}
Equation for $\beta$
\begin{equation}
	\dot{\beta} + \frac{4}{3\varphi_0 (t+t_0)} \beta + \frac32 \beta^2 -\frac{1}{(t+t_0)^2} = 0.
\end{equation}

\begin{figure}[h!]
	\center{\resizebox{!}{7.5cm}{
			\includegraphics{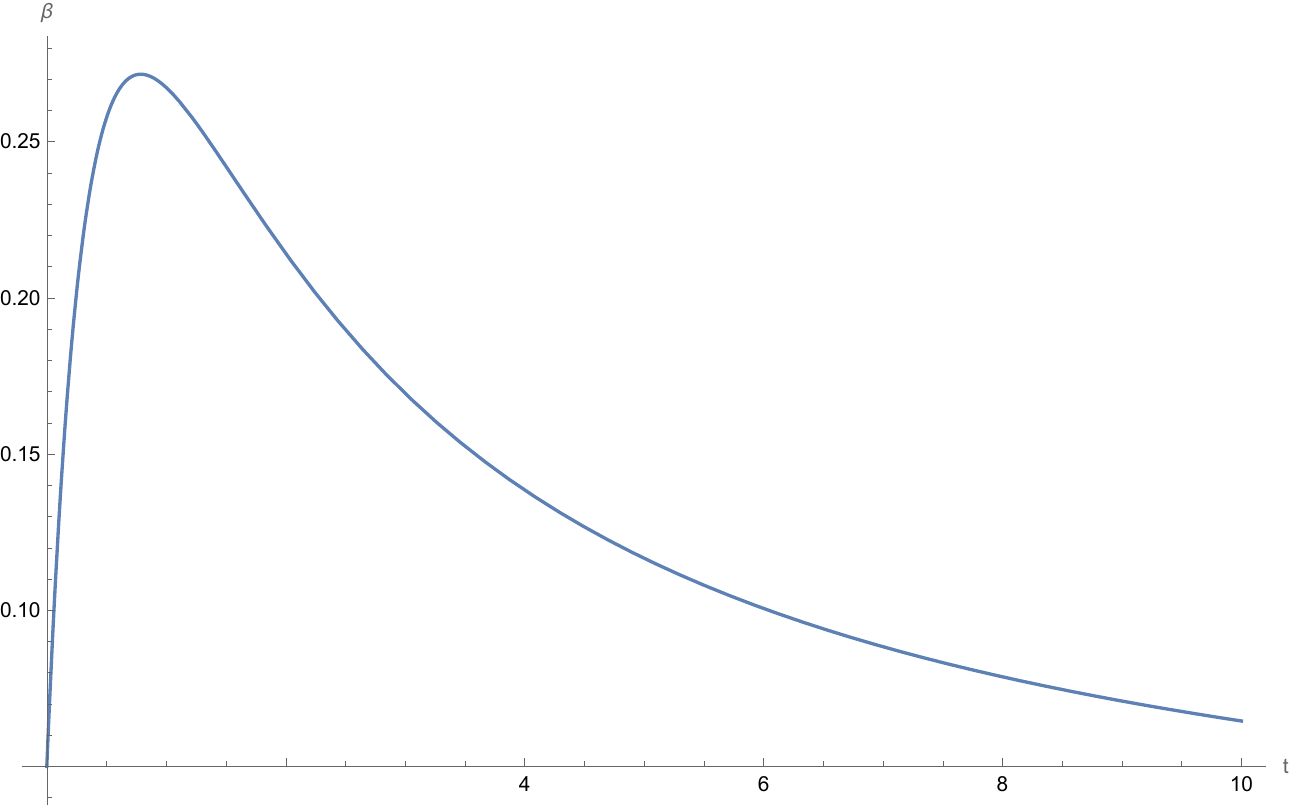}}}
	\caption{\small Lyra's parameter $\beta$ in quintessence model}\label{beta in quinessence model}
\end{figure}

\section{Conclusions}\label{sec:conclusions}

Within the framework of a Bianchi type-I cosmological model, we investigate the role of Lyra’s geometry in the evolution of the Universe when it is filled with dark energy modeled by a scalar field. Our analysis reveals that the corresponding Einstein equations retain the same form as in the absence of Lyra’s geometry. We find violation of conservation of stress-energy tensor in the presence of Lyra's geometry. From this violation we obtain dynamical equation for gauge field $\beta$. Due to nonlinearity of founded equation for the Lyra's geometry parameter, we have only numerical results for several cases. However, the dependence the parameter of Lyra’s geometry from scalar field and metric influences the final results.

In all considered cases we have the same picture: $\beta$ parameter influenced in early stages of evolution of the universe and doesn't affect in present stage. Only in case of exotic matter we obtain anomaly associated with peculiarities of scalar field in this case.

We plan to explore scalar field in LRS-BI and/or FLRW universes with Lyra’s geometry and compare the theoretical and numerical results with the observational data in upcoming papers, and we plan to construct and study correct scalar field theory within all Lyra's geometry aspects.

\end{document}